\documentclass{aa}  
\usepackage{graphicx}
\usepackage{txfonts}
\usepackage{natbib}
\newfont{\sfsl}{cmssqi8 scaled 1200}
\newfont{\sfslp}{cmssqi8 scaled 1250}
\newfont{\sfsls}{cmssqi8 scaled 900}
\newfont{\sfsln}{cmssqi8 scaled 1350}
\newfont{\sfslz}{cmssqi8 scaled 1500}
\newfont{\sfslzz}{cmssqi8 scaled 1150}
\newfont{\sfslms}{cmssqi8 scaled 1000}	
\newfont{\sfsla}{cmssqi8 scaled 3000}
\newfont{\sfslb}{cmssqi8 scaled 2200}
\newcommand{\gcss}{{\sfslms HIFLUGCS}} 
\newcommand{\gcsa}{{\sfsl HIFLUGCS}} 

\newcommand{\ro}{{\it ROSAT}}

\newcommand{\as}{{\it ASCA}}

\newcommand{\xmm}{{\it XMM-Newton}}

\newcommand{\cha}{{\it Chandra}}

\newcommand{\rosi}{{\it eROSITA}}

\newcommand{\om}{\Omega_{\rm m}}

\newcommand{\ol}{\Omega_{\Lambda}}

\newcommand{\lx}{L_{\rm X}}

\newcommand{\mt}{M_{\rm tot}}

\begin{document}
   \title{The galaxy cluster X-ray luminosity--gravitational mass relation in
   the light of the WMAP 3rd year data}

   \titlerunning{Cluster $\lx$--$\mt$ relation in light of WMAP 3rd year data}

   \author{T.~H. Reiprich}

   \offprints{T.~H. Reiprich}

   \institute{Argelander Institute for Astronomy (AIfA)\thanks{Founded by
   merging of the Institut f\"ur Astrophysik und Extraterrestrische Forschung,
   the Sternwarte, and the Radioastronomisches Institut der Universit\"at
   Bonn.}, Bonn University, 
              Auf dem H\"ugel 71, 53121 Bonn, Germany\\
              \email{thomas@reiprich.net}
             }

   \date{Received 29.04.2006; accepted 18.05.2006}
 
  \abstract
   {The 3rd year WMAP results mark a shift in best fit values of cosmological
   parameters compared to the 1st year data and the concordance cosmological
   model.}
   {We test the consistency of the new results with previous constraints on
   cosmological parameters from the \gcss\ galaxy cluster sample and the
   impact of this shift on the X-ray luminosity--gravitational mass relation.} 
   {The measured X-ray luminosity function combined with the observed
   luminosity--mass relation are compared to mass functions predicted for given
   cosmological parameter values.}
   {The luminosity function and luminosity--mass relation
   derived previously from \gcss\ are in perfect agreement with mass functions
   predicted using the best fit parameter values from the 3rd year WMAP data
   ($\om=0.238, \sigma_8=0.74$) and inconsistent with the concordance
   cosmological model 
   ($\om=0.3, \sigma_8=0.9$), assuming a flat Universe. Trying to force
   consistency with the concordance model requires artificially decreasing the
   normalization of the luminosity--mass relation by a factor of 2.}
   {The shift in best fit values for $\om$ and $\sigma_8$ has a significant
   impact on predictions of cluster abundances. The new WMAP results are now in
   perfect agreement with previous results on the $\om$--$\sigma_8$ relation
   determined from the mass function of \gcss\ clusters and other X-ray cluster
   samples (the ``low cluster normalization''). We conclude that
   -- unless the true values of $\om$ and $\sigma_8$ differ significantly from the
   3rd year WMAP results -- the luminosity--mass relation is well
   described by their previous determination from X-ray observations of
   clusters, with a conservative upper limit on the bias factor of 1.5. 
   These conclusions are currently being tested directly in a complete follow-up
   program of all \gcss\ clusters with \cha\ and \xmm.}


   \maketitle
%

\section{Introduction}
The abundance of local galaxy clusters is very sensitive to the
matter density, $\om$, and the amplitude of density fluctuations, expressed as
$\sigma_8$, in the Universe. This has long been known and the method has been
applied often to constrain cosmological parameters
\citep[e.g.,][]{ha91,bc92,bc93,pbs03,vv04}. In practice, this 
test requires detailed knowledge about the number density of clusters and their
gravitational mass. The number density as a function of, e.g., X-ray luminosity
can be measured relatively easily while the accurate determination of the
cluster mass is difficult and subject to systematic errors. 
For instance, mass determinations based on galaxy velocity dispersion and X-ray
emission of the intracluster gas may be affected if the cluster is undergoing a
merger event \citep[e.g.,][]{rsr02,rsk04}, gravitational lensing masses may be
affected by projection effects \citep[e.g.,][]{cdk04}.
In the era of precision cosmology and with the realization of very tight
constraints on cosmological parameters especially from cosmic microwave
background measurements \citep[e.g.,][]{svp03}, these difficulties have lead
to the idea to turn the argument around and assume the cosmological model to be
known precisely and use the abundance of clusters to study cluster physics.
For example, the
X-ray luminosity--gravitational mass ($\lx$--$\mt$) relation may be
\emph{derived} by combining the measured X-ray luminosity function with the
assumed cluster mass function.

In the last years, a broad consensus has developed that we live in a concordance
Universe ($\om=0.3$, $\ol=0.7$, $\sigma_8=0.9$), based on consistent
measurements from a multitude of completely independent observations (from
cosmic microwave background, distant supernovae type Ia, large scale structure,  
galaxy clusters, etc., e.g., \citealt{bop99,wtz02,t02a,t02b}); although the
$\sigma_8$ constraint remained somewhat controversial \citep[e.g.,][]{wtj03}. 
Then, the case was further strengthend when the 1st year
results from WMAP returned more evidence for a flat Universe for reasonable
values of the Hubble constant, and best fit values that are very close to the
concordance values ($\om=0.29$, $\sigma_8=0.9$, \citealt{svp03}, their Tab.~2).

\begin{figure}
\resizebox{\hsize}{!}{\includegraphics{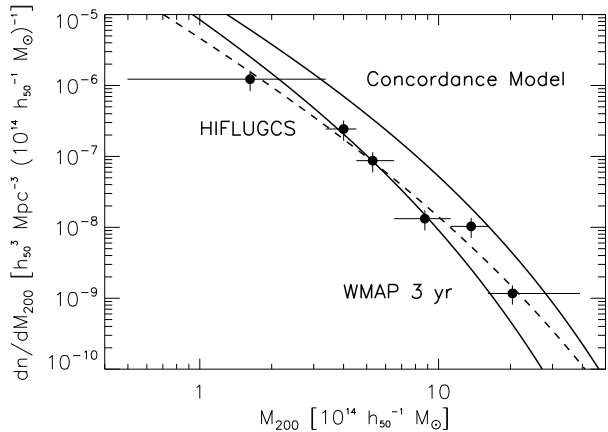}}
\caption{Galaxy cluster mass functions. Filled circles: \gcss\ data (63
clusters); dashed line: best fit to \gcss\ data; upper solid line: predicted by
concordance model ($\om=0.3,\sigma_8=0.9$), which was a very good fit to the
WMAP 1st year data; lower solid line: predicted by best fit to the WMAP 3rd year
data ($\om=0.234,\sigma_8=0.74$). The concordance model predicts too many
($>$200) clusters in the survey volume while the new cosmology predicts about
the correct number. Adapted from \citet[Fig.~10.87]{r05}.} 
\label{MF}%
\end{figure}

Although the best fit $\om$ and $\sigma_8$ values changed only
within the uncertainties between the WMAP 1st year and 3rd year data, this has
nonetheless dramatic consequences for predictions of cluster abundances, since
these are very sensitive to even small changes in these values. This is
illustrated in Fig.~\ref{MF} where the cluster mass functions for the
concordance cosmology and the new best fit values ($\om=0.238, \sigma_8=0.74$,
\citealt{sbd06}; from now on referred to as ``new cosmology'') are compared.
Also shown are data points from the \gcsa\ mass function, determined with
individual mass measurements for 63 galaxy clusters, assuming the intracluster
gas to be in hydrostatic equilibrium with the gravitational potential
\citep{rb01}. The concordance model predicts more than 200 
galaxy clusters in the survey volume while the new cosmology predicts $\sim$50
clusters. It is to be expected that this change in best fit cosmology has
significant consequences when deriving the cluster $\lx$--$\mt$ relation by
combining the measured luminosity function with the predicted mass function. The
aim of this \emph{Letter} is to illustrate these consequences and suggest a
current best estimate of the $\lx$--$\mt$ relation.
%

\section{Measuring and Predicting the Cluster $\lx$--$\mt$ Relation}
The observed cluster $\lx$--$\mt$ relation was first studied in detail by
\citet[RB02]{rb01} using the X-ray flux-limited \gcsa\ sample.
It is difficult to accurately simulate the X-ray luminosity of galaxy clusters
\citep[e.g.][]{fwb99} but it is reassuring that high resolution cosmological
hydrodynamical simulations that include cooling and non-gravitational heating
are able to reproduce this observed $\lx$--$\mt$ relation
\citep[e.g.,][]{tbs03}.

Since the scatter in this relation is finite, RB02 (Section~4.2), as
well as \citet[Section~6.4]{r01} and \citet[Section~2.3]{irb01}
mention explicitly that the normalization might be slightly biased high because,
in a flux-limited sample, clusters with higher luminosity for a given mass have a
higher likelihood to be included in the sample. In a recent detailed study,
\citet{seb06} quantify this bias by predicting the $\lx$--$\mt$
relation using cluster number counts and the scatter in the
luminosity--temperature relation, assuming the cluster mass function in a
concordance Universe ($\om=0.3$, $\ol=0.7$, $\sigma_8=0.9$). They find the
normalization measured by RB02 is biased high by a factor of 2 --
a significant bias.

\begin{figure}
\resizebox{\hsize}{!}{\includegraphics{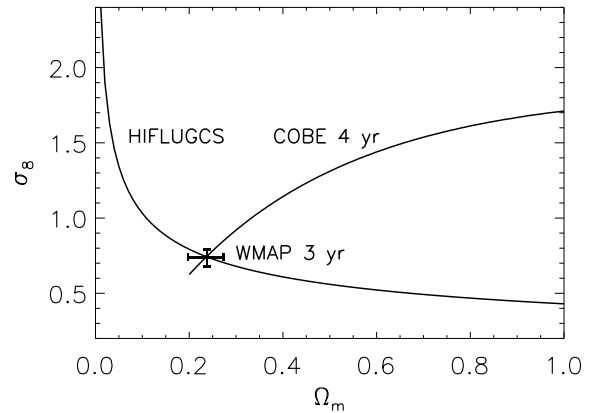}}
\caption{The best fit model to the WMAP 3rd year data lies exactly at the
intersection of the $\om$--$\sigma_8$ relations from \gcss\ and COBE. COBE
relation from \citet{bw97}, calculated using a code provided by T. Kitayama.
Adapted from \citet{r02}.} 
\label{ome-sig}%
\end{figure}

The mass function measured with \gcsa\ is not consistent with a concordance
Universe (Fig.~\ref{MF}). RB02 found the relation $\sigma_8=0.43\om^{-0.38}$,
which surprisingly implied $\sigma_8=0.68$ for a fiducial $\om=0.3$, and
discussed this ``low cluster normalization'' for the first time in
detail.
This relation is shown in Fig.~\ref{ome-sig}. Also shown are the results from
the COBE 4 year data \citep{bw97} and the WMAP 3rd year best fit value. The new
WMAP result lies exactly at the intersection of the COBE and \gcsa\ results. The
fact that the WMAP value lies exactly on the \gcsa\ curve also
implies that the new cosmology predicts the correct cluster abundance. This
suggests that the normalization of the $\lx$--$\mt$ relation measured for
the \gcsa\ clusters may actually not be biased very strongly. 
%

\begin{figure}
\resizebox{\hsize}{!}{\includegraphics{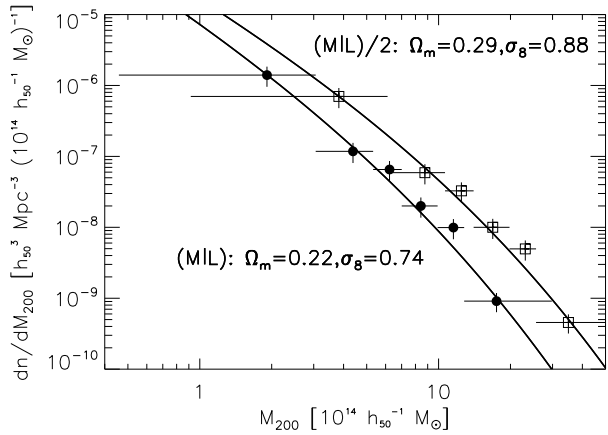}}
\caption{Quasi-mass functions estimated from luminosity function and
luminosity--mass relations. 
Filled circles: $\lx$--$\mt$ relation as measured with \gcss;
open squares: normalization of this relation artificially divided by a factor
of 2. The solid lines indicate the respective best fit mass functions:
($\om=0.22,\sigma_8=0.74$) and ($\om=0.29,\sigma_8=0.88$); the former result
is very close to the new cosmology favored by WMAP and the latter very close
to the concordance model.}
\label{MFLM}%
\end{figure}

To test this quantitatively, we determined ``quasi-mass functions'' from the
\gcsa\ luminosity function. In detail, the availability of individually measured
masses for all \gcsa\ clusters was ignored and the mass for each cluster was
instead estimated from its luminosity and the $\lx$--$\mt$ relation ($M|L$ in
Tab.~8 of RB02). With these ``masses,'' the quasi-mass function was constructed.
Then, predicted mass functions were fit to the quasi-mass function, following
the procedure described in detail in RB02, resulting in best fit 
values\footnote{Note that \citet{irb01} and RB02 demonstrated
that even for a local cluster sample; i.e., without information on cluster
abundance evolution, the degeneracy between $\om$ and $\sigma_8$ can be broken
through the shape of the mass function (the resulting direct best fit values
from the RB02 mass function are consistent within the statistical uncertainties
with the new cosmology).}
$\om=0.22$ and $\sigma_8=0.74$ -- almost precisely the values favored by the new
cosmology (Fig.~\ref{MFLM}).
Since \citet{seb06} argue that the normalization of the \gcsa\ $\lx$--$\mt$
relation is
overestimated by a factor of 2, we divided the normalization by 2 and repeated
the above procedure. The new best fit values were $\om=0.29$ and
$\sigma_8=0.88$; i.e., almost exactly the concordance cosmology assumed by
\citet{seb06}. Therefore, assuming the new cosmology from the WMAP 3rd year
data, we conclude that the factor of 2 bias found by \citet{seb06} is due to
assuming the ``wrong'' cosmological model. This conclusion is qualitatively
consistent with their relation $L_{15,0}\propto \sigma_8^{-4}$ (note that the
``$\ln$'' in the relation they give in their abstract and in their derivation in
Section~4.2  seems to be a typo), which implies a factor of 3 reduction in
normalization when going from $\sigma_8=0.68$
to $\sigma_8=0.9$.
Note that in a revised version of their paper, \citeauthor{seb06} will offer a
``compromise'' solution, assuming $\sigma_8=0.85$ (for $\om=0.24$), resulting in
moderate scatter in the $\lx$--$\mt$ relation and a smaller bias (A. Evrard,
priv.\ comm.).
%

The absence of a strong bias found here also implies that the
intrinsic scatter in the $\lx$--$\mt$ relation may indeed be relatively small.
Assume, unrealistically, that there is zero intrinsic scatter. This would imply
a bias factor of 1 (i.e., no bias at all) for cluster samples drawn from a given
mass distribution, 
regardless of how the sample was selected, because all clusters with a given
mass have exactly the same luminosity. The larger the scatter the larger the
bias in flux-limited samples, therefore, a small bias indicates small scatter.
\citet{zbf06} use 14 clusters from an almost volume-complete sample.
As mentioned in \citet{r01}, in truly volume-limited samples there should be no
bias in the normalization of the $\lx$--$\mt$ relation, regardless of the amount
of intrinsic scatter. In their
\xmm\ analysis \citeauthor{zbf06} confirm normalization and
scatter of the RB02 luminosity--mass relation, corroborating our result here
that the bias is small.

Note that we do \emph{not} argue here that the bias factor in the
normalization of the $\lx$--$\mt$ relation estimated from flux-limited samples
is exactly 1 (see our references above), but rather that a bias factor of
2 as found by \citet{seb06} is a strong overestimate unless the new WMAP results
are ``wrong'' or unless another competing systematic effect conspires to hide a
bias this large.
That the intrinsic scatter is indeed larger than zero is,
e.g., indicated by the observed segregation of cooling core clusters towards the
high luminosity side for a given mass in the $\lx$--$\mt$ relation (Y. Chen et
al., in prep.; \citealt{rs03}) and by temporary luminosity increases in major
mergers predicted by simulations \citep[e.g.,][]{rsr02}. We are currently
studying these effects in more detail with better data.

To derive an upper limit on the bias allowed by the WMAP 3rd year data we used
the largest $\om$ and $\sigma_8$ values consistent with the quoted
uncertainties: $\om=0.27$ and $\sigma_8=0.79$ \citep[note that we combined the
uncertainties in $\om h^2$ and $h$ in a worst case fashion, to be
conservative]{sbd06}. Then the quasi-mass function fit 
procedure was repeated for varying bias factors. The largest bias factor for
which the 1-$\sigma$ error ellipse is still consistent with $\om=0.27$ and
$\sigma_8=0.79$ is 1.5. Therefore, we conclude that the bias factor $\le1.5$,
unless $\om$ and/or $\sigma_8$ are larger than allowed by 
the WMAP 3rd year data. Without being less conservative, a tighter constraint on
the bias parameter seems currently not feasible with this method because of the
existing uncertainties in $\om$ and $\sigma_8$ and the strong sensitivity of the
cluster mass function on these parameters.

Despite the very good agreement with the new WMAP results, it should be noted
that in their \ro\ and \as\ analysis RB02 had to use rather simplifying
assumptions in their determination of cluster masses. For instance, the
intracluster medium was assumed to be isothermal throughout the entire cluster
volume and for some clusters only a relatively small number of photons was
available for the determination of the gas density profile. With the new
generation of X-ray satellites these limitations can be overcome. Therefore, we
are currently following-up all \gcsa\ clusters with \cha\  (Hudson et al., in
prep.; \citealt{rhe06}) and \xmm\  (Nenestyan et al., in prep.). This should
allow us to get a realistic estimate of the intrinsic scatter directly.

\section{Summary of Conclusions}

   \begin{enumerate}
      \item The shift in best fit values for $\om$ and $\sigma_8$ between WMAP
      1st year and WMAP 3rd year data has a significant effect on predictions of
      cluster abundances because they are very sensitive to even small changes
      in these parameters.
      \item The WMAP 3rd year data are now consistent with the low cluster
      normalization found previously.
      \item The observed cluster X-ray luminosity--gravitational mass relation
      and X-ray luminosity function are in perfect agreement with mass functions
      predicted for $\om=0.238$ and $\sigma_8=0.74$, as favored by the WMAP 3rd
      year data.
      \item Assuming $\om=0.238$ and $\sigma_8=0.74$, we find that the
      expected bias in normalization when determining the cluster
      luminosity--mass relation from flux-limited samples is small -- much
      smaller than the factor of 2 claimed previously.
      \item Assuming $\om=0.27$ and $\sigma_8=0.79$, the largest values allowed
      by the new WMAP data, we derive a conservative upper limit on the bias
      factor of 1.5. 
      \item A small bias indicates that the intrinsic scatter in the
      luminosity--mass relation is small.
   \end{enumerate}

The RB02 results, together with results from other works
\citep[e.g.,][]{brt01,sel01,vnl01} that yielded the low cluster normalization,
triggered a $\sigma_8$-debate, which seems to have resulted in a loss
of trust in cosmological constraints derived from clusters within the cosmology
community. Now that other well-respected methods like those that rely on
measurements of anisotropies in the cosmic microwave background also tend 
to find lower values of $\sigma_8$, we hope that trust in clusters will
be restored. Clusters are ideally suited to address the new fundamental
questions in cosmology \citep[e.g.,][]{m05}: planned X-ray cluster surveys like
those to be performed with \rosi\footnote{See, for instance,
http://www.mpe.mpg.de/erosita/MDD-6.pdf\,.} are expected to yield up to 100,000
clusters out to large redshifts. The resulting detailed measurements of the
evolution of cosmic structure are among the most promising routes to pin down
the nature and evolution of dark energy.

\begin{acknowledgements}
We thank the anonymous referee for quick and useful comments.
T.H.R. acknowledges support by the Deutsche Forschungsgemeinschaft through Emmy
Noether Research Grant RE 1462.
\end{acknowledgements}

\bibliographystyle{aa}
%

\end{document}